# Towards an Author-Topic-Term-Model Visualization of 100 Years of German Sociological Society Proceedings

Arnim Bleier and Andreas Strotmann

*{arnim.bleier, andreas.strotmann}@gesis.org*
GESIS - Leibniz Institute for the Social Sciences, Unter Sachsenhausen 6-8,
Cologne (Germany)

**Introduction**

Author co-citation studies (Zhao & Strotmann, 2008) employ factor analysis to reduce high-dimensional co-citation matrices to low dimensional and possibly interpretable factors, but these studies do not use any information from the text bodies of publications. We hypothesise that term frequencies may yield useful information for scientometric analysis. In our work we ask if word features in combination with Bayesian analysis allows for well-founded science mapping studies. This work goes back to the roots of Mosteller and Wallace's (1964) statistical text analysis using word frequency features and a Bayesian inference approach, tough with different goals. To answer our research question we (i) introduce the data set on which the experiments are carried out, (ii) describe the Bayesian model employed for inference and (iii) present first results of the analysis.

**The DGS Dataset**

The collection of documents D we use in the experiment covers ~100 years of proceedings (from 1910 to 2006) of meetings of the *Deutsche Gesellschaft für Soziologie* (DGS), a total of 5,010 documents. Early proceedings had been scanned and OCRed, others were used in original digital form. Metadata for the documents included 3,661 distinct full names of authors J.

From each document, the 21$^{st}$-320$^{th}$ words were extracted. After unifying word case, we removed stop-words, short and/or rare words (< 4 letters; > 10 occurrences; mostly OCR fragments) and words found in more than half of the documents, resulting in 1,067,128 occurrences from a vocabulary V with 12,665 distinct words.

**Statistical Model**

We now review the statistical model we employ to relate authors and documents via a flexible number of topics. Following a common notation (Rosen-Zvi, 2004), a document d is modelled as a vector of $N_d$ words, $\mathbf{w}_d$, where the i$^{th}$ word $w_{di}$ is chosen from the unique terms in vocabulary V. Each document d is associated with a set of authors $\mathbf{j}_d$ from the set of all authors, J.

Our model assumes that documents are generated in the following steps:

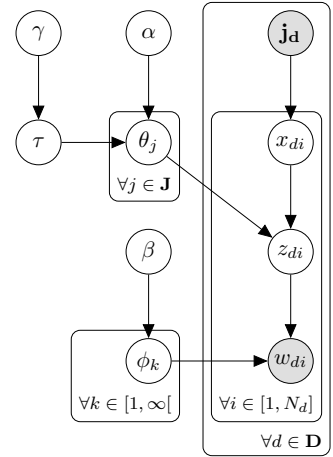

**Figure 1. Non-parametric Author-Topic model.**

1. Draw a shared discrete probability distribution from a Dirichlet Process (DP) (Teh et al. 2006) with base measure H and prior concentration parameter $\gamma$ as a global mixture over topics
$$\tau \sim DP(H, \gamma)$$
2. For author j, draw an author specific distribution over topics from the global topic mixture $\tau$, with prior concentration parameter $\alpha$
$$\theta_j \sim DP(\tau, \alpha)$$
3. For each topic k, draw a topic specific distribution over vocabulary V from the symmetric Dirichlet prior $\beta$
$$\phi_k \sim Dir(\beta)$$
4. For all $N_d$ words in document d, (i) draw an author indicator x from the set of authors $\mathbf{j}_d$ of document d; (ii) a topic indicator z from the author specific topic distribution $\theta_j$; (iii) the observed word w itself from the respective topic
$$x_{di} \sim Discrete(\mathbf{j}_d)$$
$$z_{di} \sim Discrete(\theta_{x_{di}})$$
$$w_{di} \sim Discrete(\phi_{z_{id}})$$

Figure 1 illustrates the independence assumptions made by the generative storyline via plate notation. Circles represent statistical variables, with observed ones shaded. Arrows represent conditional dependence − i.e., the order in which variables are drawn. Plates indicate repetition, as indicated by universal quantifiers.

For the posterior analysis, the topic distributions $\phi_k$ over terms as well as the author distributions $\theta_j$

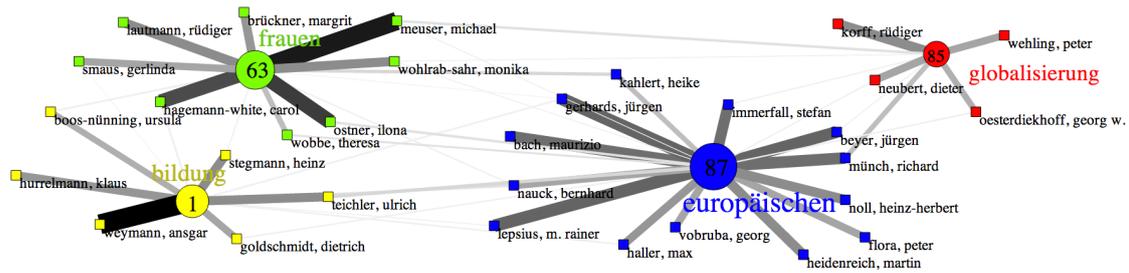

**Figure 1**: Excerpt of author topic analysis result: visualization of four topics and their main authors

over topics are of particular interest. The former span a latent semantic space via meaningful word probabilities for each topic; the latter allow us to position each author j in this topic space.

**Posterior Analysis and Visualization**

The generative model is structured as a directed acyclic graph beginning from causes and ending with observed words in documents. Bayes' Rule reverses causality, and parameters of interest can be estimated from observed data and priors. We use an MCMC sampler for this posterior analysis. After running the sampler for 2,000 steps with priors $\gamma=.5$, $\alpha=.5$ and $\beta=.2$, the model converged to 89 components. Samples of $\phi_k$ and $\theta_j$ are shown in Figure 1 and Table 1. The reader is referred to Rosen-Zvi et al. (2004), Teh et al. (2006) and Bleier (2012) for an in-depth discussion of this method.

**Table 1. Most probably words for the topics.**

| Topic | High probability words |
|---|---|
| 1 | bildung, jugendlichen, schule, jugendliche, ausbildung, jugendlicher <br> **translation**: education, school, youth |
| 63 | frauen, männern, männer, geschlecht, frauenforschung, frau <br> **translation**: women, men, gender research |
| 85 | globalisierung, welt, globalen, grenzen, globaler, unternehmen <br> **translation**: globalization, world, enterprize |
| 87 | europäischen, europa, integration, europäische, union, ländern <br> **translation**: European, integration, countries |

Due to space constraints we restrict the visualization in Figure 1 (using Pajek) to four of 89 components. Authors and topics are represented as square and circular nodes, resp. The size of topic nodes is proportional to their usage and the strength of the arcs proportional to $\theta_{jk}$, the probability for topic k specific to author j. We are not constrained to interpreting topics as only having a distinct probability for each author, but equally have for each topic k a distribution $\phi_k$ over distinct words in the vocabulary. Table 1 displays the six most probable terms for the sample topics of Figure 1.

**Discussion**

Our approach to science mapping uses a flexible version latent Dirichlet allocation to (i) identify an optimal number and set of topics for a given set of documents based on the words that occur in them, (ii) to identify the most relevant words to describe each topic, and (iii) to identify weighted links between authors and the topics of their writings. The statistical model takes into account that documents are written by multiple authors, that authors write on different topics to different degrees, and that words pertain to different topics to varying degrees.

Figure 1 shows a small fragment of a map of German sociological science based on ~100 years of DGS proceedings, inspired by the visualization of results of co-citation-based factor analysis in Zhao & Strotmann (2008), but generated fully automatically from the results of applying this statistical analysis technique to full texts.

While a full evaluation remains to be done, these results show some promise for the application of these methods in scientometric studies.